\documentclass[%
aps,prd,onecolumn,showpacs,showkeys,amsmath,amssymb,superscriptaddress,floatfix,nofootinbib
,notitlepage]{revtex4-1}

\usepackage{graphicx}
\usepackage{dcolumn}
\usepackage{bm}
\usepackage[utf8]{inputenc}

\usepackage{color}
\usepackage{ulem}

\newcommand{\RNum}[1]{\uppercase\expandafter{\romannumeral #1\relax}}

\begin{document}

\preprint{APS/123-QED}

\title{Asymptotic Safety in the Conformal Hidden Sector?}
\author{Zhi-Wei Wang}
\author{Frederick S.~Sage}
\author{T.G.~Steele}
\affiliation{
Department of Physics and
Engineering Physics, University of Saskatchewan, Saskatoon, SK,
S7N 5E2, Canada
}
\author{R.B.~Mann}
\affiliation{
Department of Physics, University of Waterloo, Waterloo, ON, N2L 3G1, Canada
}

\begin{abstract}
We combine the notion of asymptotic safety (AS) with conformal invariance in a hidden sector beyond the Standard Model. We use the renormalization group (RG) equations as a bridge to connect UV boundary conditions and EW/TeV scale physics and furnish a detailed example in the context of a leptophobic $U(1)'$ model. 
A broad selection of UV boundary conditions  are formulated corresponding to differing AS scenarios, and we find  an AS scenario with very strong predictive power, allowing unique determination of most of the parameters in the model. We obtain the interrelationships among the couplings, the transition scale of the fixed point $M_{UV}$ and the generations of quarks coupled to the $Z'$, and especially the correlation between $M_{UV}$ and the top quark Yukawa coupling $Y_t$. 
 Several phenomenological implications of our results are presented for selected $Z'$ masses.   
\end{abstract} 
\maketitle

\section{Introduction}
 
 The Standard Model (SM) faces both the hierarchy/naturalness problem and the triviality problem \cite{Callaway:1988ya}. According to Bardeen's insight, conformal symmetry as a custodial symmetry can protect the Higgs mass from large UV contributions \cite{Bardeen}, addressing the naturalness problem \cite{'tHooft:1979bh}. Such a possibility can be realized if the conformal symmetry is only softly broken, 
 requiring the couplings to run to a fixed point at a high energy scale \cite{Tavares:2013dga} and thus the UV sensitivity is under control \cite{Abel:2013mya}. On the other hand, the Standard Model (SM) itself is not UV complete since the $U(1)$ hypercharge gauge coupling and Higgs quartic coupling will reach a Landau Pole in the far UV. The notion of asymptotic safety (AS) was proposed to make the SM and gravity valid up to arbitrarily high energy scales without any singularities \cite{Shaposhnikov:2009pv,Holthausen:2011aa}. In this article, we make a connection between the fixed point required in the (quantum) conformal scenario to address the hierarchy/naturalness problem and the notion of asymptotic safety (AS) to address the UV completeness of the SM, which also requires a fixed point \cite{Shaposhnikov:2009pv,Holthausen:2011aa} and thus both problems may be solved simultaneously. Since it is highly non-trivial to   realize a fixed point within the SM, the requirement of a fixed point provides both an excellent motivation for extending the SM and  a very strong constraint on low energy physics.




In this article, we 
implement the AS principle in hidden sector models  with classical conformal symmetry and apply this formalism to the detailed example of a leptophobic $U(1)^\prime$ model, and explore some phenomenological applications. We emphasize that only quantum conformal symmetry can address the Higgs hierarchy/naturalness problem, and not  classical conformal symmetry. However, in this investigation we also require that the scalar sector satisfy classical conformal symmetry in order to study the possibility of radiative symmetry breaking via the Coleman-Weinberg (CW) mechanism~\cite{Coleman:1973jx}. The CW mechanism \cite{Coleman:1973jx} has the advantage of dynamically generating natural scale hierarchies between the unification scale and the electroweak (EW) scale, through dimensional transmutation as in QCD \cite{Weinberg:1978ym,Hill:2014mqa} and has strong predictive power.  Although different versions of SM extensions with classical conformal symmetry have been proposed in the literature \cite{Hempfling:1996ht,Altmannshofer:2014vra}, none have incorporated AS. We stress that it is highly non-trivial to realize the CW mechanism and asymptotic safety simultaneously when gravity is involved.  As we will later demonstrate, the scalar quartic coupling $\lambda$ must run to a Gaussian fixed point  to be UV safe when gravity is involved,  implying $\beta_\lambda<0$ on the RG flow, whereas the CW mechanism requires $\beta_\lambda>0$. Here we find a very intriguing solution: the quartic coupling can have $\beta_\lambda>0$ at low energies,  satisfying the requirement of the CW mechanism  and $\beta_\lambda<0$ at high energies, realizing UV safety.

 We regard asymptotic safety as providing deep motivation for the existing non-AS flat scenarios proposed in \cite{Hashimoto:2013hta}, with an emphasis on the fixed points of the scalar couplings below the singularity to avoid Landau poles and provide a stable Higgs vacuum. Introducing the model in section II, we provide a categorization of different AS scenarios according to the gravity contribution to the RG functions above Planck scale. We use the RG equation as a bridge connecting UV boundary conditions to EW/TeV scale physics and explore the implications for SM observables.
The predictive power of AS scenarios and further constraints from the requirement of realizing symmetry breaking via the CW mechanism imply that most of the parameters in the model are uniquely determined, thereby providing interesting interrelationships amongst the couplings,  the scale of the fixed point (transition scale) $M_{UV}$, and the generations of quarks coupled to the $U(1)^\prime$ gauge field. 
Furthermore, it is not necessary to associate the  the transition scale $M_{UV}$ with the Planck scale; indeed we find that $M_{UV}$ is very sensitive to the top quark Yukawa coupling $Y_t$, and for a fixed $Y_t$ there exists an interesting range of values for $M_{UV}$.  The details of the RG equation analysis are in section III.
Section IV contains an application of these techniques to 
 a variety of observables at the LHC. 
Our results are summarized in section V.


\section{Model Building}

We investigate the classical conformally symmetric complex singlet extension of the SM with an extra $U(1)'$ gauge symmetry. The Lagrangian of the scalar sector is written as
\begin{equation}
L=D_\mu H^{\dagger}D^\mu H+D_\mu S^{\dagger} D^\mu S-\lambda_2\left|S\right|^2 H^{\dagger}H
-\lambda_3\left|S\right|^4-\lambda_1\left(H^{\dagger}H\right)^2,
\end{equation}
where $H$ is the SM Higgs doublet, $S$ is the complex singlet and $D_\mu$ is the extended covariant derivative. In the basis where the two $U(1)$ gauge kinetic terms are diagonal, the covariant derivative term is written as \cite{Hashimoto:2013hta,Chankowski:2006jk}
\begin{equation}
D_\mu=\partial_\mu-ig_3\frac{\lambda_a}{2}G_\mu^a-ig_2\frac{\tau_i}{2}W_\mu^i-iY\left(g_YB_\mu+g_{m}B_\mu'\right)-ig'Q_B'B_\mu',
\end{equation}
where $g_3$, $g_2$, $g_Y$ and $g'$ are the gauge couplings of $SU(3)_c$, $SU(2)$, $U(1)_Y$ and $U(1)'$ respectively. The quantities $Y$ and $Q_B'$ denote the $U(1)_Y$ hypercharge and the $U(1)'$ charge.  We make explicit the mixing term proportional to $g_m$ that couples the $B_\mu^\prime$ field to SM hypercharge $Y$. Dilepton constraints on new neutral gauge bosons are stringent \cite{Aad:2014cka}, so we would like to avoid coupling the $U(1)^\prime$ gauge group to SM leptons at tree level, which we do by making the model leptophobic. Because we would like to focus on a leptophobic variant, we choose a special case of the gauge group $U(1)_{B-xL}'$ where $x=0$ and the gauge group in our case can be denoted $U(1)_B'$ with charge $Q_B'$ \cite{Carena:2004xs}. Our analysis has found that the fixed points, as discussed in the next section, are highly sensitive to the number of fermions in the theory. To obtain realistic solutions to the renormalization group equations, we have chosen to set the $U(1)_B'$ charge of the first and second generations of SM quarks to zero. This choice will be discussed further in subsequent sections. For convenience, we list the charge assignments of the SM fermion and scalar sectors of the $U(1)_B'$ model in Tables \ref{fermion gauge charge} and \ref{Scharge}.  Several other sets of charge assignments that avoid anomalies can be found in Ref.~\cite{Carena:2004xs}.

Inclusion of a new gauge group generates a new set of perturbative anomalies. To cancel these anomalies, we include several particles charged under the new gauge group $U(1)_B'$. In particular, we require a right handed neutrino $\nu_R$ and two `spectator' fermions $\psi^l_{L,R}$ and $\psi^e_{L,R}$. Charge assignments for these exotic fermions and the neutrino are included in Table~\ref{exotic fermion gauge charge}. 

Since our model does include a right handed neutrino, we briefly comment on some aspects of the neutrino sector.
The neutrino couples to the singlet field through a Majorana Yukawa-type term \cite{Carena:2004xs,Hashimoto:2013hta}: $L_M=-Y_M^{ij}\bar{\nu_{Ri}^c}\nu_{Rj}S+\left(\rm{h.c.}\right)$. We have left the flavor indices explicit, though our model only requires a single right handed neutrino to match the single generation of quarks. The neutrino will acquire a Majorana mass of $m_{\nu_R}=\sqrt{2}v_1Y_M$ after $U(1)_B$ symmetry breaking, with $v_1$ the vacuum expectation value of the scalar field.  The model, as  presented, cannot generate light neutrino masses. However, by adding a Higgs-neutrino Yukawa term \cite{Hashimoto:2013hta}
\begin{equation}
L_D=-Y_D^{ij}\bar{\nu_{Ri}}H^\dagger \l_{Lj}+\left(\rm{h.c.}\right)
\end{equation}
a Dirac neutrino mass term can be generated. In combination with the heavy Majorana mass term above, the seesaw mechanism \cite{Langacker:2008yv, Iso:2009nw} can be enabled, explaining the light neutrino mass hierarchy. Actually, we expect the Higgs-neutrino Yukawa coupling $Y_D^{ij}$ to be small and asymptotically free and so it will not change the results we obtain.
We leave such extensions of our model for future work, and refer the interested reader to discussions in many similar situations \cite{Hashimoto:2013hta, Iso:2009nw}.

The spectator fermions $\psi^{l,e}_{L,R}$ are vectorlike under the SM gauge group. The masses of these particles are assumed to be much larger than the $\rm{TeV}$ scale.  
The inclusion of any of the possible Yukawa terms (Dirac or Majorana) between the $\psi^{l,e}_{L,R}$ and the scalars (either the Higgs $H$ or the singlet $S$) would violate one of the gauge symmetries present;  for that reason, the $\psi^{l,e}_{L,R}$ do not interact with the scalars at tree level. These spectator particles have no effect on our phenomenological conclusions. The mass of these vector-like fermions can be introduced directly through explicitly mass terms. Since they do not directly couple to the scalar fields, they have no contributions to the Higgs mass and will not reintroduce the naturalness issue. In addition, 
these explicit mass terms will not spoil the CW mechanism, since it only requires the scalar sector (rather than the whole system) to be classically scale invariant.
\begin{center}
\begin{table}
\caption{Standard Model fermion charge assignments}
\label{fermion gauge charge}
\begin{tabular}{|c||c|c|c|c|c|c|c|c|c|c|c|}
\hline
Gauge group & $q^{u,d}_L$ & $u_R$  & $d_R$ & $q^{c,s}_L$ & $c_R$  & $s_R$ & $q^{t,b}_L$ & $t_R$  & $b_R$ & $l_L$ (all gens) & $e_R$ (all gens)  \\
\hline\hline
 $SU(3)_c$ & 3 & 3 & 3 & 3 & 3 & 3 & 3 & 3 & 3 & 1 & 1 \\
\hline
 $SU(2)_L$ & 2 & 1 & 1 & 2 & 1 & 1 & 2 & 1 & 1 & 2 & 1 \\
\hline
 $U(1)_Y$ & 1/3 & 4/3 & -2/3 & 1/3 & 4/3 & -2/3 & 1/3 & 4/3 & -2/3 & -1 & -2 \\
\hline
 $U(1)^\prime$ & 0 & 0 & 0 & 0 & 0 & 0 & 1/3 & 1/3 & 1/3 & 0 & 0 \\
\hline
\end{tabular}
\end{table}
\end{center}

\begin{center}
\begin{table}
\caption{Scalar charge assignments}
\label{Scharge}
\begin{tabular}{|c||c|c|}
\hline
 Gauge group & $S$ & $H$  \\
\hline\hline
 $SU(3)_c$ & 1 & 1  \\
\hline
 $SU(2)_L$ & 1 & 2 \\
\hline
 $U(1)_Y$ & 0 & 1/2 \\
\hline
 $U(1)^\prime$ & 2 & 0 \\
\hline
\end{tabular}
\end{table}
\end{center}

\begin{center}
\begin{table}
\caption{Exotic fermion charge assignments}
\label{exotic fermion gauge charge}
\begin{tabular}{|c||c|c|c|c|c|c|}
\hline
  Gauge group & $\nu_R$ &$\psi^l_L$ &$\psi^l_R$ &$\psi^e_L$ &$\psi^e_R$  \\
\hline\hline
 $SU(3)_c$ & 1 &1 & 1 & 1 & 1 \\
\hline
 $SU(2)_L$ & 1 & 2 & 2 & 1 & 1 \\
\hline
 $U(1)_Y$ & 0 & -1 & -1 & -2 & -2 \\
\hline
 $U(1)^\prime$ & -1 & -1 & 0 & -1 & 0 \\
\hline
\end{tabular}
\end{table}
\end{center}

\section{Renormalization Group Analysis}

We consider in this paper the gravity contributions to the SM and Beyond SM (BSM) sector by treating gravity as an effective theory where at low energy scale the theory should match Einstein-Hilbert gravity. In order that the theory not break classical conformal symmetry and be compatible with the classically conformal particle physics sector, we also assume a general fundamental gravity theory that possesses a classical conformal symmetry at  the UV scale and at a low energy scale dynamically generates the Einstein-Hilbert term through a symmetry-breaking mechanism (induced gravity) \cite{Zee:1978wi,Adler:1982ri} (see  e.g., Refs.~\cite{Holdom:2015kbf,Einhorn:2014gfa,Salvio:2014soa} for more recent work).
 A sample classically conformally symmetric gravity model is given in \cite{Einhorn:2014gfa,Salvio:2014soa,Shapiro:1989dq,Accioly:2016qeb}:
\begin{equation}
S = \int d^4x \sqrt{|\det g|} \,\bigg[ \frac{R^2}{6f_0^2} + \frac{\frac13 R^2 -  R_{\mu\nu}^2}{f_2^2} - \xi_H |H|^2 R -\xi_S |S|^2 R+ 
L^{\rm conf}_{\rm SM}+L^{\rm conf}_{\rm BSM}\label{agravity_model}
\bigg]\,,
\end{equation}
where $L^{\rm conf}_{\rm SM}\,,L^{\rm conf}_{\rm BSM}$ correspond respectively to the classically conformal version of the SM Lagrangian and beyond SM Lagrangian (i.e.~without mass term).
This theory has the advantage of being renormalizable and asymptotically free \cite{Stelle:1976gc,Fradkin:1981iu}
 It is clear that in this sample model, the Einstein-Hilbert term at low energy scale can be dynamically generated after symmetry breaking (either perturbatively or non-perturbatively) and $S$ obtains a vacuum expectation value. Ghost issues in scale-invariant higher-derivative gravity theories have been addressed  in Refs.~\cite{Holdom:2015kbf,Einhorn:2014gfa,Donoghue:2018izj} and also in non-scale-invariant theories \cite{Modesto:2015ozb,Accioly:2016qeb}.

We henceforth focus only on the gravity contribution to the SM/BSM system and do not consider the feedback to the gravity system. As we shall see, only the signs of the gravity contributions will be important in our later analysis rather than their detailed form. Thus it is sufficient to treat  gravity as an effective field theory without using the full power of a fundamental theory as in Eq.~\eqref{agravity_model}. The leading order gravity contribution to the SM/BSM couplings have the following form (see \cite{Shaposhnikov:2009pv} for more details):
\begin{equation}
\beta_j^{tot}=\beta_j^{\rm{SM/BSM}}+\beta_j^{\rm{grav}},\quad\beta_j^{\rm{grav}}=\frac{a_j}{8\pi}\frac{k^2}{M_p^2(k)}x_j\,,\label{gravity RG}
\end{equation}
where $a_j$ is a coefficient whose sign will be very important later on and $k^2$, $M_p^2(k)$, $x_j$ are the energy scale, Planck mass and  corresponding coupling respectively.  We shall neglect the running of the gravitational coupling because $a_j<<1$ below $M_{UV}$, and so $M_{UV}$ is well below the scale where gravity plays an important role.
The form of the gravity contribution can be understood in an effective theory context since the effective gravitational coupling scales as $k/M_p$.
It should be emphasized that the gravity contribution to the RG function of a certain coupling is actually proportional to the coupling itself. 
It has been shown that the gravity contribution to the SM RG functions is negative i.e.~$\beta_j^{\rm{grav}}<0$ for the gauge couplings (gauge couplings are realized as asymptotically free even for $U(1)$ hypercharge) \cite{Robinson:2005fj,Shaposhnikov:2009pv} as well as the Yukawa couplings \cite{Shapiro:1989dq,Shaposhnikov:2009pv}.
Thus, all SM gauge couplings and top Yukawa coupling are asymptotically free and valid to arbitrarily high energy scales. However, for the SM Higgs quartic coupling, the leading order gravity contribution is positive i.e.,~$\beta_{\lambda_1}^{\rm{grav}}>0\,\left(a_\lambda\sim3.1\right)$ \cite{Percacci:2003jz,Narain:2009fy}. Note that the calculations shown in \cite{Percacci:2003jz,Narain:2009fy} are for a general scalar quartic coupling and thus the leading order gravity contribution is also positive for the singlet quartic coupling i.e.~$\beta_{\lambda_3}^{\rm{grav}}>0$.
Thus, to realize the asymptotic safety scenario, we require the Higgs quartic coupling $\lambda_1$ to reach a Gaussian fixed point i.e.,~$\lambda_1\left(M_{UV}\right)=0$, $\beta_{\lambda_1}\left(M_{UV}\right)=0$.  

Note that the Higgs quartic coupling cannot reach an interacting fixed point (i.e.~at 
$\lambda\neq0$)  because the gravity contribution $\beta_\lambda^{\rm{grav}}=\frac{a_\lambda}{8\pi}\frac{k^2}{M_p^2\left(k\right)}\lambda$ in Eq.~\eqref{gravity RG}  would not vanish,  
 therefore spoiling the fixed point. 
In this scenario, the whole system is UV complete and all the SM couplings reach a Gaussian fixed point around the Planck scale.
 Note also that the UV boundary condition provides a stable Higgs vacuum directly.

In the $U(1)'_B$ complex singlet extension of the SM, the extra gauge coupling $g'$ will be taken care of by gravity for the same reason discussed above \cite{Robinson:2005fj,Shaposhnikov:2009pv} and becomes asymptotically free in the UV.
The boundary conditions at the transition scale can be categorized according to the gravity contribution to the singlet quartic and Higgs portal running couplings $\lambda_3$ and $\lambda_2$ respectively. For $\beta_{\lambda_2}^{\rm{grav}}>0$, this implies the fixed point conditions $\lambda_2\left(M_{UV}\right)=0, \; \beta_{\lambda_2}\left(M_{UV}\right)=0$; however (as we shall see below) $\beta_{\lambda_2}\left(M_{UV}\right)=0$ is not phenomenologically viable and thus the case $\beta_{\lambda_2}^{\rm{grav}} > 0$ is ruled out.
We therefore only consider $\beta_{\lambda_2}^{\rm{grav}}<0$ ($\beta_{\lambda_2}^{\rm{grav}}=0$ will lead to UV incompleteness of $\lambda_2$). Although $\lambda_2$ consequently need not run into a fixed point at $M_{UV}$, it might still be of interest to consider $\lambda_2\left(M_{UV}\right)=0$ as one of the conditions where the Higgs portal interaction is purely radiatively generated. 

It has been shown in \cite{Percacci:2003jz,Narain:2009fy} that the leading order gravity contribution to a general scalar quartic coupling is positive and thus $\beta_{\lambda_3}^{\rm{grav}}>0$.
We therefore focus on the following UV boundary condition corresponding to AS and quantum conformal symmetry
 \begin{equation}
\beta_{\lambda_1}\left(M_{UV}\right)=\beta_{\lambda_3}\left(M_{UV}\right)=\lambda_1\left(M_{UV}\right)=\lambda_2\left(M_{UV}\right)=\lambda_3\left(M_{UV}\right)=0\label{boundary1}
\end{equation}
where the UV scale $M_{UV}$ is not necessarily the Planck scale $M_{pl}$.
 The boundary condition \eqref{boundary1} is also particularly interesting because it has the strongest predictive power of the three scenarios, determining all parameters in the model.  However, if the gravity contributions are highly non-perturbative,  (going  beyond the existing perturbative calculation or even functional RG analysis \cite{Percacci:2003jz,Narain:2009fy} ), it might still be possible to realize the case $\beta_{\lambda_3}^{\rm{grav}}<0$ with the boundary conditions
\begin{align}
\beta_{\lambda_1}\left(M_{UV}\right)&=\lambda_1\left(M_{UV}\right)=0; \quad\lambda_2\left(M_{UV}\right),\, \lambda_3\left(M_{UV}\right)\neq0\label{boundary2}\\
\beta_{\lambda_1}\left(M_{UV}\right)&=\lambda_1\left(M_{UV}\right)=\lambda_2\left(M_{UV}\right)=0; \quad\lambda_3\left(M_{UV}\right)~\neq0\label{boundary3}
\end{align}
 which allow greater freedom in the parameter space of the model.  However, based on the evidence of Refs.~\cite{Percacci:2003jz,Narain:2009fy} we shall focus on the case \eqref{boundary1}.

Electroweak symmetry breaking can be realized  through two  sequential  steps: (i) through the CW mechanism \cite{Coleman:1973jx}, radiative symmetry breaking first occurs in the singlet hidden sector at energy scale $v_1$ higher than the EW scale and (ii) is then communicated to the Higgs sector to trigger EW symmetry breaking with a small Higgs portal interaction (inversely proportional to $v_1$). However, it is highly non-trivial to realize the Coleman-Weinberg mechanism in the singlet hidden sector in scenario \eqref{boundary1}. For the boundary condition $\beta_{\lambda_3}\left(M_{UV}\right)=\lambda_3\left(M_{UV}\right)=0$ in scenario \eqref{boundary1}, it normally implies that $\beta_{\lambda_3}<0$ (similar to asymptotically free theory) on the RG flow while to realize the Coleman-Weinberg mechanism it normally requires $\beta_{\lambda_3}>0$  because the coupling should run from positive to negative from UV to IR to trigger symmetry breaking. However, we find an intriguing solution (explicated below) that $\beta_{\lambda_3}<0$ at a high energy scale while $\beta_{\lambda_3}>0$ at a lower energy scale, which guarantees that the singlet quartic coupling reaches a Gaussian fixed point in the UV and triggers radiative symmetry breaking in the IR through the CW mechanism.


We are able to study these two scalar sectors separately. Letting $H=\frac{1}{\sqrt{2}}\left(\phi_1+i\phi_2, \phi_3+i\phi_4\right)$, $S=\frac{1}{\sqrt{2}}\left(\varphi_1+i\varphi_2\right)$ and defining $\phi^2=\sum_{i}\phi_i^2$ and $\varphi^2=\sum_{i}\varphi_i^2$, we obtain the vacuum expectation value (VEV) condition of the Higgs sector $\frac{dV_{Higgs}}{d\phi}|_{\phi=v\atop \varphi=v_1}=0$,
where $V_{Higgs}=\frac{1}{4}\lambda _1 \phi^4+\frac{1}{4}\lambda _2\phi^2\varphi^2$ is the Higgs effective potential and $v,v_1$ are identified with the electroweak scale $v=246\,\rm{GeV}$ and the singlet breaking scale $v_1$ respectively.
We only use the tree level Higgs effective potential because conventional symmetry breaking is completely determined by the tree level VEV condition. Combining the Higgs sector VEV condition with the $125\,\rm{GeV}$ Higgs mass observed \cite{2012gk,2012gu}, we obtain
\begin{equation}
2v^2\lambda_1=-v_1^2\lambda_2=M_H^2~,\label{VEV & Mass}
\end{equation}
where $M_H=125\,\rm{GeV}$ is the Higgs mass and $\lambda_1,\,\lambda_2$ are evaluated at the EW scale.  

To realize  radiative symmetry breaking in the singlet sector, we consider the one loop RG improvement of the singlet effective potential
\begin{equation}
V_{S}=\frac{1}{4}\lambda_3\left(t\right)G^4\left(t\right)\varphi^4=\frac{1}{4}\lambda_3\left(t\right)G^4\left(t\right)v_1^4{\exp\left(t\right)}^4~,\label{V_S expression}
\end{equation}
where $t\equiv\log\left(\varphi/v_1\right)$ with the renormalization scale at the VEV of the singlet, $G\left(t\right)\equiv\exp\left[-\int_0^t dt'\gamma\left(t'\right)\right]$ and $\lambda_3\left(t\right)$ is the running singlet self-coupling. The VEV condition of the singlet is defined by
\begin{equation}
\frac{dV}{d\varphi}\bigg|_{\varphi=v_1}=\left(\frac{e^{-t}}{v_1}\right)\frac{dV_{\rm{S}}}{dt}\bigg|_{t=0}=0~,\label{VEV2 MS}
\end{equation}
where t=0 corresponds to the singlet broken scale $v_1$. 
The RG functions are obtained using the formula in \cite{Hashimoto:2013hta} with charge assignments in Table \ref{fermion gauge charge} and the assumption that $Z'$ only couples to  one generation of quarks (discussed below) while the anomalous dimension is provided in \cite{Iso:2009nw}. At one loop level for the beyond-SM part and three loop level for the SM part, they are written as:
\begin{equation}
\begin{split}
16\pi^2\beta_{\lambda_1}&=\lambda_2^2-3\lambda_1g_{m}^2+\frac{3}{8}g_{m}^4+\beta^{SM}_{\lambda_1}\\
16\pi^2\beta_{\lambda_2}&=12g_{m}^2g'^2+6Y_t^2\lambda_2-24g'^2\lambda_2+4Y_M^2\lambda_2+4\lambda_2^2+12\lambda_1\lambda_2+8\lambda_2\lambda_3-\frac{3}{2}\lambda_2\left(g_{m}^2+3g_2^2+g_1^2\right)\\
16\pi^2\beta_{\lambda_3}&=96g'^4-16Y_M^4+2\lambda_2^2-\lambda_3\left(48g'^2+8Y_M^2-20\lambda_3\right)\\
16\pi^2\beta_{Y_t}&=-\frac{17}{12}Y_tg_{m}^2-\frac{2}{3}Y_tg'^2-\frac{5}{3}Y_tg'g_{m}+\beta^{SM}_{Y_t}\\
16\pi^2\beta_{g'}&=\frac{1}{18}g'\left(76g'^2+64g'g_{m}+123g_{m}^2\right)\\
16\pi^2\beta_{g_{m}}&=g_{m}\left(\frac{41}{6}\left(g_{m}^2+2g_1^2\right)+\frac{38}{9}g'^2\right)+\frac{32}{3}g'\left(g_{m}^2+g_1^2\right)\\
16\pi^2\beta_{Y_{M}}&=-6Y_Mg'^2+6Y_M^3;~~32\pi^2\gamma_\varphi=Y_M^2-24g'^2\label{RG}
\end{split}
\end{equation}
where $\beta^{SM}_{\lambda_1}$ and $\beta^{SM}_{Y_t}$ are the three loop SM RG functions provided in \cite{Buttazzo:2013uya}.  
Three-loop SM RG functions are necessary because higher-loop contributions from the top quark Yukawa coupling $Y_t$ are too large to be ignored, while the effects of hidden-sector  couplings are small enough to be well-approximated by the one-loop RG functions. 
The $Z'$ mass 
\begin{equation}
M_{Z'}=2g'\left(0\right)v_1
\label{Z' mass}
\end{equation}
provides a further constraint. Our analysis, while quite general, is best illustrated by choosing three specific samples:~$M_{Z'}=1.9,\,3.4,\,6.8\,\rm{TeV}$.
If future experiments see evidence for different $Z'$ candidates, Eq.~\eqref{Z' mass} could alter our inputs accordingly. We also note that  $\beta_{\lambda_2}\left(M_{UV}\right)=0$ is not consistent with  \eqref{Z' mass}. From Eq.~\eqref{RG}, $\beta_{\lambda_2}\left(M_{UV}\right)=0$ implies $g_m\left(M_{UV}\right)g'\left(M_{UV}\right)=0$. 
If $g'\left(M_{UV}\right)=0$, $g'$ will be negative below the scale $M_{UV}$ since $\beta_{g'}>0$,  while if $g_m\left(M_{UV}\right)=0$, the coupling $\lambda_2$ will run very slowly from $M_{UV}$ to the EW scale, leading to a $\lambda_2$ too small at the EW scale  to satisify \eqref{VEV & Mass} and \eqref{Z' mass}, and justifying our choice to consider only $\beta_{\lambda_2}^{\rm{grav}}<0$.

We now solve the RG equations with the $UV$ boundary conditions \eqref{boundary1} using the RG functions \eqref{RG}, reducing the system of
nine unknowns $\left(\lambda_1,\,\lambda_2,\,\lambda_3,\,g',\,g_{m},\,Y_M,\,Y_t,\,v_1,\,M_{UV}\right)$ to that of five ($g_{m}$, $v_1$, $Y_M$, $Y_t$, $M_{UV}$) by employing the constraints  \eqref{VEV & Mass}, \eqref{VEV2 MS}, \eqref{Z' mass}.  
Our solutions are shown in Table \ref{BC}. Note that these solutions are evaluated at the $U(1)'$ broken scale and the SM top Yukawa coupling needs to run to the EW scale to compare with the measured value.
Interestingly, we find $M_{UV}\sim M_{pl}$ and the predicted $Y_t=0.93$ to be very close to the current experimental central value $Y_t^c=0.936$ \cite{Buttazzo:2013uya}.

\begin{figure}[htb]
\centering
\includegraphics[width=0.6\columnwidth]{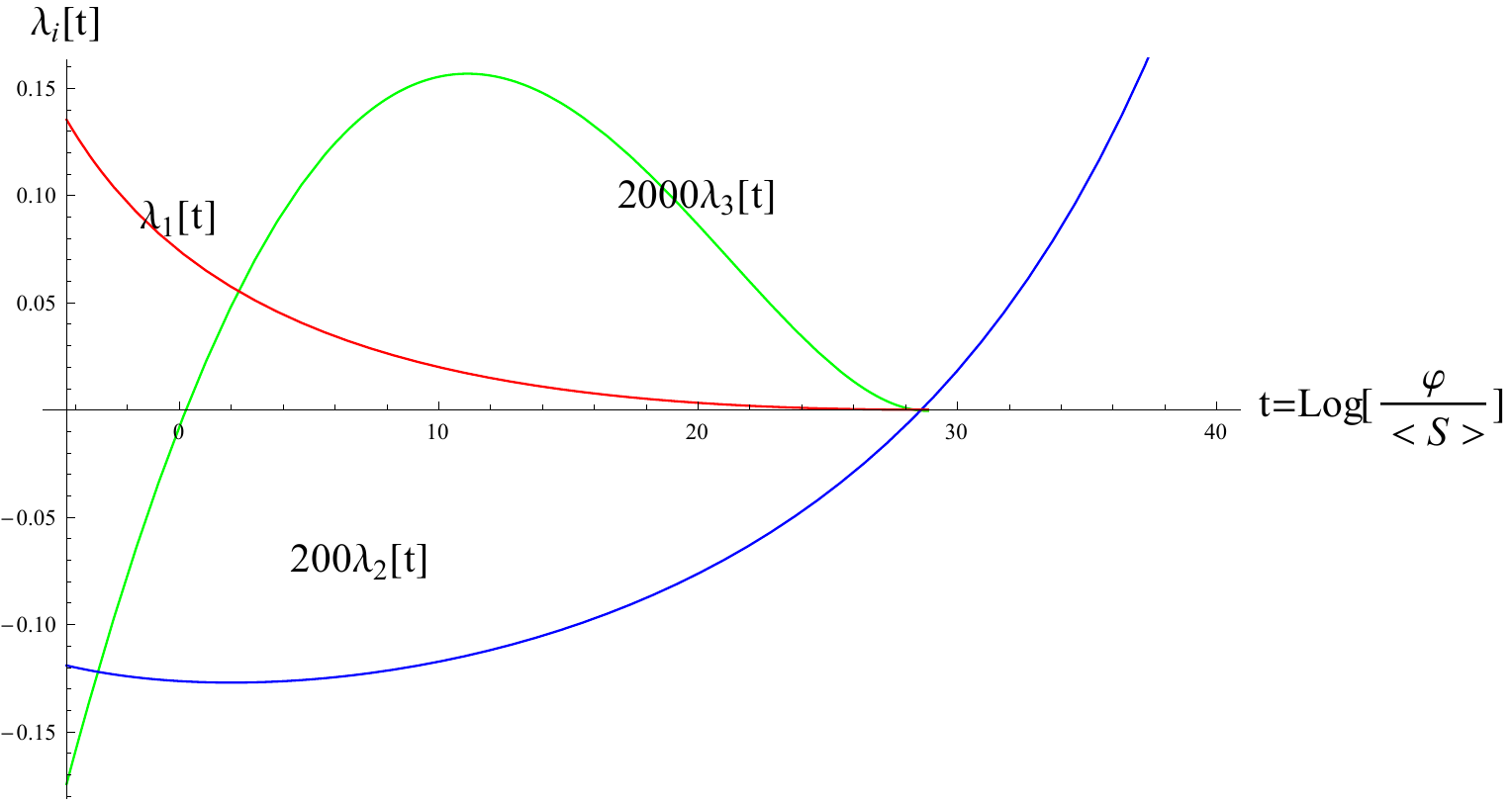}\hspace{0.04\columnwidth}
\caption{Running scalar couplings are shown as a function of the scale $t=\log\left(\varphi/\langle S\rangle\right)$. The red, blue, and green curves represent $\lambda_1\left(t\right)$, $200\lambda_2\left(t\right)$, $2000\lambda_3\left(t\right)$ respectively. }
\label{running coupling3}
\end{figure} 
We plot the running scalar couplings from the EW scale to the Planck scale in Figure \ref{running coupling3}.  
To satisfy the boundary condition $\beta_{\lambda_3}\left(M_{UV}\right)=\lambda_3\left(M_{UV}\right)=0$ in scenario \eqref{boundary1}, $Z'$ can only couple to at most two 
generations of quarks, a point also emphasized in \cite{Hashimoto:2013hta} for the $U\left(1\right)_{B-L}$ model. This is because $\beta_{g'}$  increases  when the  $Z'$ couples to more generations of quarks. 
 A large $\beta_{g'}$ leads to the fast running of $\lambda_3$ and $\beta_{\lambda_3}>0$ throughout the RG flow until a Landau Pole is encountered, which is not consistent with the slow running of $\lambda_3$ at UV (i.e.~$\beta_{\lambda_3}<0$ at UV) required to satisfy the boundary conditions \eqref{boundary1}, as illustrated in Figure \ref{running coupling3}. Because of this constraint and an additional phenomenological constraint (flavour changing neutral currents) that will be discussed in the next section, we assume $Z'$ only couples to the third generation of quarks.

The singlet quartic coupling $\lambda_3$, from UV to IR, runs from positive to negative (i.e.~$\beta_{\lambda_3}>0$ at IR) satisfying the CW mechanism requirement. The transition point where $\lambda_3$ runs to zero defines the singlet breaking scale $v_1$ by the CW mechanism \cite{Coleman:1973jx}, a point also emphasized in \cite{Altmannshofer:2014vra}. It might seem surprising that the singlet quartic coupling can run to negative values. However, the true effective coupling defined as $\lambda_{eff}=\frac{e^{-4t}}{v_1^4}\frac{d^4V_{\rm{s}}}{dt^4}|_{t=0}$ provides a positive effective coupling value of $\lambda_{eff}\left(0\right)=1.2\times 10^{-4}$. In addition, we are able to show that the singlet mass predictions in both MS scheme and CW scheme are different by 2\%, implying the consistency of both schemes.

 The results of our analysis are presented in Table \ref{BC}. 
\begin{table}[hbt]
\centering
\begin{tabular}{||c|c|c|c|c|c|c|c|c|c|}
\hline
$M_{Z^\prime}$ (TeV) &$10^6\lambda_2$ & $10^6\lambda_3$ & $g'$ & $g_{m}$ & $Y_M$ & $Y_t$ & $m_{\nu_{R}}$ (TeV) & $v_1$ (TeV) & $m_S$ (GeV) \\ 
\hline 
1.9 & -6 & -4.05 & 0.18 & 0.042 & 0.28 & 0.77 & 2.02 & 5.1 & 20 \\ 
\hline 
3.4 & -0.59 & -0.13 & 0.1 & 0.023 & 0.16 & 0.74 & 3.72 & 16.9 & 11.5 \\ 
\hline 
6.8 & -0.036 & -0.002 & 0.05 & 0.011 & 0.08 & 0.70 & 7.59 & 68.8 & 6 \\ 
\hline 
\end{tabular}
\caption{This table summarizes our solutions for the $\overline{\rm MS}$ scheme couplings (evaluated at the $U(1)'$ broken scale $\sim v_1$) and the VEV of the singlet $v_1$ in TeV units according to the boundary condition  (BC) \eqref{boundary1}  for different choices of $M_{Z'}$.}
\label{BC}
\end{table}


\section{Collider Phenomenology}
The introduction of a new $Z^\prime$ boson that interacts with SM fermions necessitates an examination of  possible collider constraints and signatures. In this section, we consider constraints on our model from the 8 TeV LHC data, as well as making predictions for an array of signatures at the 13 and the 14 TeV LHC. 
 Because the AS scenario requires a  $Z^\prime$ coupling to one generation, these
non-universal couplings can introduce flavor changing neutral currents (FCNCs) due to the necessary transformation between the gauge and mass eigenstates. Constraints on FCNCs that mix the first and second generations of SM fermions are very strong, but constraints on models that couple only to the third generation are considerably weaker. Following the analysis of reference \cite{Frampton:1996bs}, which shows the strongest constraint comes from neutral charmonium mixing and using the updated value of the $D^0$ mass difference \cite{Olive:2016xmw}, we find that our model satisfies experimental constraints if the $Z'$ couples to only the third generation.

In the narrow resonance approximation, the differential cross section for production of a $Z^\prime$ boson of rapidity $y$ from the collision of two protons is \cite{Langacker:2008yv}
\begin{equation}
\frac{d}{dy}\sigma_{pp\rightarrow Z^\prime} = \frac{4 \pi^2 x_1 x_2}{3 m_{Z^\prime}^3 } \sum_i \left[ f^A_{q_i}(x_1)f^B_{\bar{q}_i}(x_2) + f^A_{\bar{q}_i}(x_1)f^B_{q_i}(x_2) \right] \times \Gamma_{Z^\prime \rightarrow \bar{q}_i q_i }.
\end{equation}
The Bjorken scaling variables $x_1$ and $x_2$ are related to the rapidity through 
\begin{equation}
x_{1,2} = \frac{m_Z^\prime}{\sqrt{s}}e^{\pm y}
\end{equation}
where the center of mass energy of the proton collision is $2E = \sqrt{s}$. The functions $\{f^A_{q_i}(x_1), f^B_{\bar{q}_i}(x_2) \}$ are the proton parton distribution functions. In our calculations we have used the NLO MSTW grids \cite{Martin:2009iq}. The cross section is obtained after integration with respect to rapidity over the region $-\ln ( \sqrt{s} / m_{Z^\prime} ) \le y \le \ln ( \sqrt{s} / m_{Z^\prime} )$.
The cross section for production of a $Z^\prime$ in proton-proton collisions can then be written
\begin{equation}\label{production}
\sigma\left(pp\rightarrow Z'\right)\simeq C_s
\left(\frac{2\Gamma\left(Z'\rightarrow u\bar{u}\right)+\Gamma\left(Z'\rightarrow d\bar{d}\right)}{\rm{GeV}}\right)
\end{equation}
The computed values for $C_s$ at the LHC energies and $Z^\prime$ mass values of interest are presented in table \ref{Cs}.

\begin{center}
\begin{table}
\caption{$Z^\prime$ production cross section numerical coefficents $C_s$ (units of 1/GeV$^3$)}
\label{Cs}
\begin{tabular}{|c||c|c|c|}
\hline
 $M_{Z^\prime}$ & $\sqrt{s} = 8$ TeV & $\sqrt{s} = 13$ TeV & $\sqrt{s} = 14$ TeV  \\
\hline\hline
 1.9 TeV & 2.14 $\times 10^{-10}$ & 9.32 $\times 10^{-10}$ & 1.11 $\times 10^{-9}$ \\
\hline
 3.4 TeV & 1.52 $\times 10^{-12}$ & 2.69 $\times 10^{-11}$ & 3.64 $\times 10^{-11}$ \\
\hline
 6.8 TeV & 2.80 $\times 10^{-20}$ & 1.83 $\times 10^{-14}$ & 4.55 $\times 10^{-14}$ \\
\hline
\end{tabular}
\end{table}
\end{center}

The decay width of the $Z^\prime$ to fermions is given by \cite{Hisano:2015gna}
\begin{equation}
\Gamma_{Z^\prime\rightarrow \bar{f}f} = \Theta \left( m_{Z^\prime} - 2m_f\right)\frac{N_c m_{Z^\prime}}{12\pi}\sqrt{1-\frac{4m_f^2}{m_{Z^\prime}^2}}\left[v_f^2 \left(1+\frac{2 m_f^2}{m_{Z^\prime}^2}\right) + a_f^2 \left(1-\frac{4 m_f^2}{m_{Z^\prime}^2}\right) \right].
\end{equation}
This result holds for all fermions to which the $Z^\prime$ can decay, including SM fermions and right handed neutrinos $\nu_R$ when the masses of the latter are small enough. The vector $v_f$ and axial $a_f$ charges of the fermions can be written in terms of the SM hypercharges $Y_{iL,R}$ and the $U(1)^\prime$ charges $Q^\prime_{iL,R}$ of the left and right handed components as recorded in Table \ref{exotic fermion gauge charge}, as well as the $U(1)^\prime$ gauge coupling $g^\prime$ and the mixing parameter $g_m$:
\begin{equation}\label{AV}
v_f = g^\prime \left( \frac{Q^\prime_{fL} + Q^\prime_{fR}}{2}\right) + g_m \left( \frac{Y_{fL} + Y_{fR}}{2}\right) \qquad 
a_f = -g^\prime \left( \frac{Q^\prime_{fL} - Q^\prime_{fR}}{2}\right) - g_m \left( \frac{Y_{fL} - Y_{fR}}{2}\right). 
\end{equation} 
We have assumed that the spectator fermions $\psi$ are far too massive to be a kinematically allowed decay channel for the $Z^\prime$.

The small mass of the singlet required by some of these scenarios allows for the possibility that the SM Higgs may acquire a new decay channel $h \rightarrow SS$. As bounds on the invisible Higgs decay width are stringent, we must test whether these mass and coupling values for the singlet are compatible with LHC observations. Fortunately, the very small interaction between the singlet and the Higgs makes the model consistent with ATLAS and CMS bounds \cite{ATLASinvHiggs, CMSinvHiggs}.

We note that because the $Z^\prime$ couples dominantly to the third generation of quarks, it is possible to produce it via gluon fusion through a heavy quark loop, similarly to the SM Higgs. However this production mechanism is suppressed by the inverse of the heavy quark mass $m_Q$ squared. Moreover, the $Z^\prime gg$ interaction is proportional only to the axial charge of the quark Eq.~\eqref{AV}, and the part of that quantity that is proportional to $g^\prime$ is zero. The gluon fusion contribution is thus proportional to $g_m/m_Q^2$ and so is subdominant to the contribution from light quark annihilation. For this reason it is ignored.

The collider signatures in the narrow resonance approximation $\sigma\left(pp\rightarrow Z'\right)\times\rm{BR}\left(Z'\rightarrow final\right)$ of this model for the three $Z^\prime$ mass values have been computed for a variety of channels and presented in Table \ref{predicts} in units of femtobarns (fb). Only the $\sqrt{s}=8$ TeV case can be compared against ATLAS and CMS results, as the analyses for higher center of mass energies have not yet been reported. As well, the $M_{Z^\prime}=$ $6.8\,\rm{TeV}$ case is effectively beyond the reach of the 8 TeV LHC, and so is also unconstrained. We have considered the dilepton \cite{Aad:2014cka}, dijet \cite{Khachatryan:2015sja}, and ditop \cite{Khachatryan:2015sma} channels.
\begin{center}
\begin{table}
\caption{$Z^\prime$ predicted production cross sections $\sigma_{pp\rightarrow Z^\prime} \times BR$ (units of fb)}
\label{predicts}
\begin{tabular}{|c||c||c|c|c|}
\hline
 $M_{Z^\prime}$ & Channel & $\sqrt{s} = 8$ TeV & $\sqrt{s} = 13$ TeV & $\sqrt{s} = 14$ TeV  \\
\hline\hline
 1.9 TeV & Diboson & 0.013 & 0.058 & 0.069 \\
 & Dilepton & 4.27 & 18.59 & 22.14 \\
 & Dijet & 19.47 & 84.79 & 100.98 \\
 & Ditop & 5.67 & 24.71 & 29.42 \\
\hline
 3.8 TeV & Diboson & 9.15 $\times 10^{-5}$ & 1.62 $\times 10^{-3}$ & 2.19 $\times 10^{-3}$ \\
& Dilepton & 0.029 & 0.518 & 0.701 \\
& Dijet & 0.110 & 1.95 & 2.64 \\
& Ditop & 0.028 & 0.491 & 0.665 \\
\hline
 6.8 TeV & Diboson & 7.50 $\times 10^{-13}$ & 4.90 $\times 10^{-7}$ & 1.22 $\times 10^{-6}$ \\
& Dilepton & 2.40 $\times 10^{-10}$ & 1.57 $\times 10^{-4}$ & 3.90 $\times 10^{-4}$ \\
& Dijet & 9.41 $\times 10^{-10}$ & 6.14 $\times 10^{-4}$ & 1.53 $\times 10^{-3}$ \\
& Ditop & 2.48 $\times 10^{-10}$ & 1.62 $\times 10^{-4}$ & 4.03 $\times 10^{-4}$ \\
\hline
\end{tabular}
\end{table}
\end{center}

From comparison with ATLAS and CMS results, the diplepton constraints reported in \cite{Aad:2014cka} rule out the 1.9 TeV $Z^\prime$. 
The $3.4\,\rm{TeV}$
 $Z^\prime$ is consistent with observed constraints in all channels, and the $6.8\,\rm{TeV}$ $Z^\prime$ is unconstrained by the 8 TeV LHC, as mentioned above. Our numerical results are presented with a perturbative higher order QCD $K$ factor of 1, though our conclusions hold even with a conservative $K$ factor of 1.5. The numerical values for the 13 and 14 TeV LHC are included as predictions of our model. A basic comparison indicates that the dijet channel is much larger, due to the inclusion of the bottom quark, which has $U(1)^\prime$ charge. The dilepton signal is about 25\% of the dijet signal for $\sqrt{s}$ = 13 TeV and about 20\% for $\sqrt{s}$ = 14 TeV. However, the dilepton signal is typically far cleaner than the dijet signal, with current dilepton constraints \cite{Aad:2014cka} an order of magnitude stronger than dijet constraints. Because of this, the dilepton signal is the primary phenomenological signature of this model, despite the fact that it is nominally leptophobic. The ditop signature turns out to be not as important as the dijet and dilepton signatures, but because the model preferentially couples to the third generation, it is important to consider.

Another interesting signature of $Z^\prime$ models is the diboson signature 
see e.g., \cite{Aad:2015owa,Khachatryan:2015sja,Brehmer:2015cia,Hisano:2015gna,Bezrukov:2014ina}. 
For the diboson decay mode, we have \cite{Hisano:2015gna}
\begin{equation}
\Gamma\left(Z'\rightarrow WW\right)=\frac{g_{m}^2}{48\pi}Y_H^2M_{Z'},\label{decay to W}
\end{equation}
where $Y_H$ is the hypercharge of the Higgs field. The mixing angle of $Z$ and $Z'$ is strongly constrained by electroweak precision measurements. In our leptophobic model, this constraint is $\sin\theta\leq0.008$ \cite{Erler:2009jh} 
and our coupling solutions are in comfortable agreement with the constraint. The diboson production rates for our scenarios are included in Table \ref{predicts}. These rates are very small for the 8 TeV LHC, as expected, but become potentially observable for the 14 TeV LHC.

\section{Discussion and Conclusions}
In this work, we addressed the Naturalness/Hierarchy problem and the origin of symmetry breaking within a gravity motived UV complete model that at the same time provides interesting phenomenological signatures at the IR scale.

We have made a connection between the fixed points required in the (quantum) conformal and  asymptotic safety scenarios to study an extension of the Standard Model that is asymptotically safe with a conformally symmetric hidden sector. This connection provides a possibility that two of the deepest questions in particle physics -- i.e.~Naturalness/ Hierarchy and UV completion --  may deeply connected to each other and can be solved simultaneously.

To further address the origin of the electroweak symmetry breaking, we showed that a Coleman-Weinberg mechanism is compatible with a gravity motivated UV complete theory featuring a UV fixed point. Electroweak symmetry breaking is realized sequentially in our theory. The singlet field $S$ obtains a VEV through dimensional transmutation and is further transmitted to the Higgs sector to trigger electroweak symmetry breaking through the Higgs portal interaction.

To connect with  phenomenology, we  employed a particular $U(1)'$ leptophobic model as a practical realization of this idea.
Table \ref{BC} categorizes the   AS scenario  we explored  for different values of $M_{Z'}$. The strong predictive power of the AS safety scenario defined by the UV boundary conditions \eqref{boundary1} allows determination of most of the parameters in the model and allows us to find their interrelationships, especially the correlations among $M_{UV}$ and $Y_t$.  
The realization of Coleman-Weinberg mechanism further constrains the singlet self-coupling $\lambda_3$ and $g'$.
To realize these solutions and satisfy FCNC constraints, the $Z'$ can only couple to the third generation. 

Phenomenological consequences of our model's predictions were explored for LHC observables.
In the  range of  $M_{Z'}$ that we explored, our model satisfies current experimental constraints for two of the three sample mass values considered
and makes predictions that could be observed relevant to the future LHC program. Our work paves the way for future model construction and phenomenological study of asymptotically safe scenarios.

\acknowledgements
Z.-W.Wang thanks W.\,Xue and F.\,A.\,Chishtie for helpful discussion. T.G.S.\ and R.B.M.\ are grateful for financial support from the Natural Sciences and Engineering Research
Council of Canada (NSERC).


\end{document}